\documentclass{PoS}
\usepackage{siunitx}
\usepackage{todonotes}
\usepackage{placeins}
\usepackage{graphicx}
\usepackage{caption}
\usepackage{subcaption}
\usepackage{tikz}

\title{IceAct, small Imaging Air Cherenkov Telescopes for IceCube}

\ShortTitle{IceAct}

\author{
The IceCube Collaboration\footnote{For collaboration list, see PoS(ICRC2019) 1177.}\\
{\itshape \href{http://icecube.wisc.edu/collaboration/authors/icrc19_icecube}{http://icecube.wisc.edu/collaboration/authors/icrc19\_icecube}}\\
E-mail: \email{merlin.schaufel@icecube.wisc.edu, karen.andeen@icecube.wisc.edu}
}

\abstract{
IceAct is a proposed surface array of cost effective and compact Silicon Photomultipliers (SiPM) based small-size (50 cm) Imaging Air Cherenkov Telescopes above the IceCube in-ice detector. In coincidence with the in-ice and surface components of IceCube it forms a hybrid detector that enables new measurements combining the information from the Cherenkov light image, the surface particle footprint and the in-ice muon tracks of extensive air showers. During January 2019, two new versions of the IceAct telescope demonstrators featuring 61 SiPM pixels and improved optics were installed in the center of the IceTop surface detector at the geographic South Pole. Combining information from these two telescopes and IceCube, it is possible to test the performance in primary particle discrimination, detector calibration, and veto capabilities. We present the status of the project and the prospects of the upcoming data taking season during the Antarctic winter.
\\

\vspace{4mm}
{\bfseries Corresponding authors:}
Merlin Schaufel$^{1}$, \speaker{Karen Andeen}$^{2}$\\
{$^{1}$ \itshape RWTH Aachen University}\\
{$^{2}$ \itshape Marquette University}
}

\FullConference{36th International Cosmic Ray Conference -ICRC2019-\\
		July 24th - August 1st, 2019\\
		Madison, WI, U.S.A.}

\begin{document}

\section{Introduction}

High-energy neutrinos are a unique probe to study the extreme high-energy universe. Neutrinos reach us from their production sites in the universe without absorption or deflection by magnetic fields.
The IceCube Neutrino Observatory has discovered a flux of high-energy neutrinos of cosmic origin
\cite{Aartsen:2014gkd,Aartsen:2016xlq}. The observed neutrino flux arrives almost isotropically at Earth. Recently, evidence for correlated neutrino and photon emission from the active galaxy TXS 0506+056 has been reported \cite{IceCube:2018dnn,IceCube:2018cha}.

IceCube's main instrument \cite{Aartsen:2016nxy} is a large-volume Cherenkov detector that instruments the glacial ice at the South Pole between depths from \SIrange{1.45}{2.45}{km} with \num{5160} digital optical modules
(DOMs) each containing a \SI{10}{"} photomultiplier tube \cite{Abbasi:2010vc} and associated electronics \cite{Abbasi:2008aa}. These DOMs are frozen into the ice along 86 vertical strings, with 60 DOMs per string.
In addition to the in-ice detector, the surface is instrumented with the IceTop air-shower detector \cite{IceCube:2012nn} that is composed of 162 Cherenkov tanks, each containing about 3 cubic meters of clear ice, instrumented with two DOMs.

\begin{figure}[htb]
	\centering
	\begin{subfigure}[b]{0.47\textwidth}
		\centering
		\includegraphics[width=\textwidth]{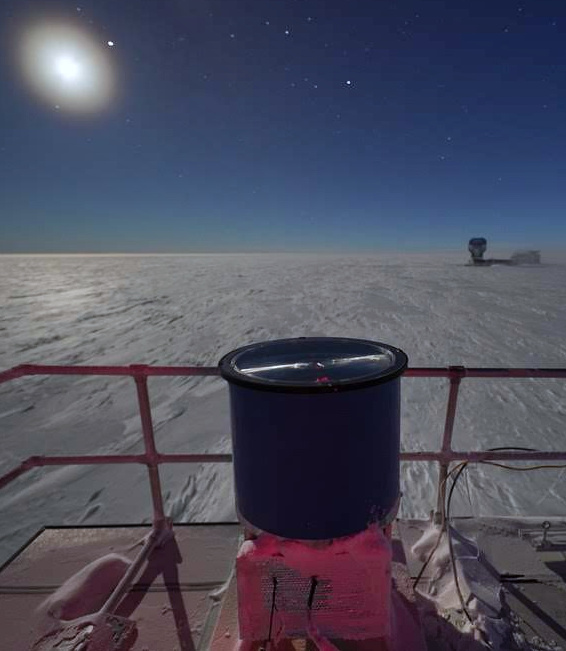}
		\subcaption{IceAct roof-telescope on the roof of the IceCube Laboratory (ICL) in the antartic winter 2019. }\label{Pic:RoofACT}
	\end{subfigure}
	\hfill
	\begin{subfigure}[b]{0.47\textwidth}
		\centering		
		\includegraphics[width=\textwidth]{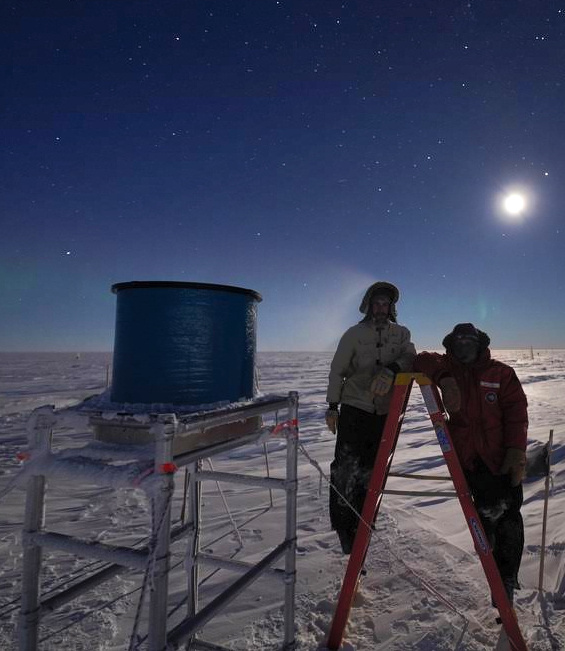}
		\subcaption{IceAct field-telescope on the snow during the antartic winter 2019.}\label{Pic:SnowACT}
	\end{subfigure}
	\caption{IceAct demonstrator telescopes operating in 2019. Images courtesy of Benjamin Eberhardt.}
\end{figure}

IceAct itself is the proposed array of multiple IceAct telescopes on the footprint of IceCube as a surface extension. Currently, it consists of two demonstrator telescopes, one telescope on the roof of the IceCube Laboratory (ICL) (see picture \ref{Pic:RoofACT}) and one on the snow approximately 220\,m west-southwest from the ICL (see picture \ref{Pic:SnowACT}). The main goal of IceAct is the calibration of the IceCube in-ice reconstruction and the IceTop energy scale as well as cosmic-ray physics in the energy range between the knee and the ankle, roughly from \SIrange{e15}{e18}{eV}.

\section{Science scope}

As a result of the observations with IceCube,
the IceCube-Gen2 collaboration aims to substantially enhance the sensitivity of IceCube for astrophysical neutrino measurements \cite{Aartsen:2014njl,Aartsen:2015dkp}.
Three detector systems have been proposed to enhance the surface detector IceTop \cite{Haungs:2019ylq} as potential extensions: a dense array of scintillator detectors %\cite{Collaboration:2017tdy}
\cite{TheIceCubeGenCollaborationHuberKelleyKunwar2017_1000082468}, radio antennas \cite{V.:2017kbm} and an array of small air-Cherenkov telescopes (IceAct) that is subject of this paper. The main goals of IceAct overlap to varying degrees with those of the other proposed extensions, and include calibration of IceCube and IceTop, measurement of the cosmic ray composition in coincidence with IceCube and IceTop, and veto of high energy muon neutrinos.  \\

\textbf{Calibration:} The coincident detection of cosmic ray-induced air showers and muons deep in the ice will allow for an improved calibration of the in-ice detector and IceTop. Direct measurements of the electromagnetic component of the air-shower with imaging air-Cherenkov telescopes can be compared to the high-energy muon component measured deep in the ice and the mixed component in IceTop. The independence of the telescopes from the ice and snow properties potentially provides a handle to reduce the influence of these systematic uncertainties  \cite{Auffenberg:2017vwc,Ackermann:2017pja}. The granularity of the camera of imaging air Cherenkov telescopes allows for precise measurements of cosmic ray-induced showers with very few telescopes.\\

\textbf{Composition Measurement:}
The observation of comic rays through several independent detection channels will also improve the capabilities of IceCube, IceTop, and IceAct to measure the composition of cosmic rays. High energy gamma-ray detection might also be possible \cite{Auffenberg:2017ypn}. A limited number of telescope stations to cover the overlap region of IceTop and IceCube in-ice promises a cost-efficient way to add an independent component to improve composition measurements in the energy range of IceCube in-ice and IceTop.\\

\textbf{Neutrino Veto:} The ability to veto high-energy muon events detected with IceCube, when a surface detector detects a coincident air-shower signal, will reduce the background to astrophysical neutrino searches and lower the detection threshold \cite{Auffenberg:2017vwc,Ackermann:2017pja}. The low energy threshold might greatly increase the sensitivity of IceCube in the southern sky for astrophysical neutrino detection down to \SI{30}{TeV} neutrino energy or lower with a dedicated array of IceAct telescopes \cite{Auffenberg:2017vwc}.

\section{Telescope design}

The basic concept of the IceAct imaging air-Cherenkov telescopes is a compact and robust design, as outlined in \cite{Bretz:2018lhg}, optimized for operation in extreme environments and cost efficiency. Thus, the telescope has enclosed optics with a large field-of-view. This enclosure shields all delicate instrumentation (including the SiPM-based camera) from the harsh South Pole environment.  Due to the use of SiPMs for the camera and the enclosure of the entire electronics, this instrument allows for a high duty cycle as demonstrated by the FACT telescope \cite{Knoetig:2013tja,2017NIMPA.876...17N}.
%like the camera based on SiPMs. 
%The application of SiPMs in Cherenkov telescope was demonstrated successfully in the First G-PAD Cherenkov (FACT) telescope \cite{2013JInst...8P6008A,2014JInst...9P0012B}. 
The IceAct telescopes are much smaller than most imaging air-Cherenkov telescopes (IACTs), with a diameter of \SI{55}{cm} and a tube length of about \SI{1}{m} including the DAQ. This reduction in size implies a higher energy threshold (\SI{10}{TeV} to \SI{20}{TeV} primary energy) than other IACTs. However, the energy threshold for air-shower detection of imaging air-Cherenkov telescopes is naturally much lower than achievable for sparse particle detectors on the surface as the air-Cherenkov light of air showers is predominantly emitted by the electromagnetic part of the air-shower during the entire air shower development.  Thus, the energy threshold of IceAct is still much lower than that of IceTop, so this does not impede the capability of IceAct to achieve its science goals.  Compared to previous non-imaging air Cherenkov arrays at the South Pole, like VULCAN \cite{DICKINSON2000114}, imaging air-Cherenkov telescopes can monitor the entire evolution of the particle cascade propagating through the atmosphere.  In addition, the smaller field-of-view of the single pixels of the IceAct camera allows for a more selective trigger which reduces the vulnerability to fluctuations in the night-sky background light (such as auroras that distribute their light over large regions of the sky) which were problematic for VULCAN. 

\section{First light}
\FloatBarrier

\begin{figure}
	\centering
	\includegraphics[width=0.55\textwidth]{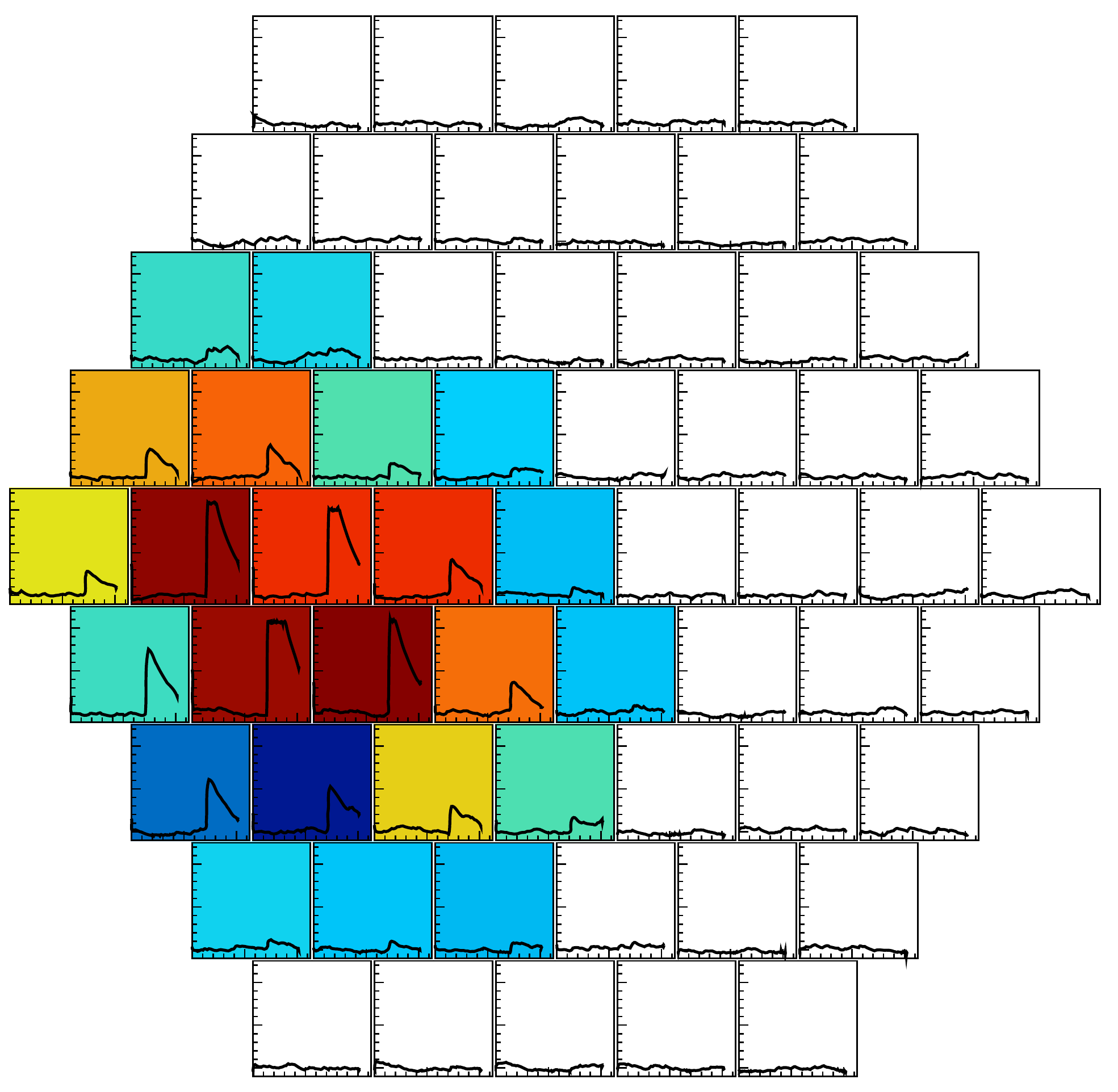}
	\caption{\label{Plt:FirstLightDRS}Event recorded shortly after the First Light of the IceAct roof-telescope (DRS DAQ) during the Antarctic winter 2019. The color encodes the value of the amplitude. The trace of every single channel (0-2000\,mV, 2\,GS/s) is also shown. }
\end{figure}

\begin{figure}
	\centering
	\includegraphics[width=0.99\textwidth]{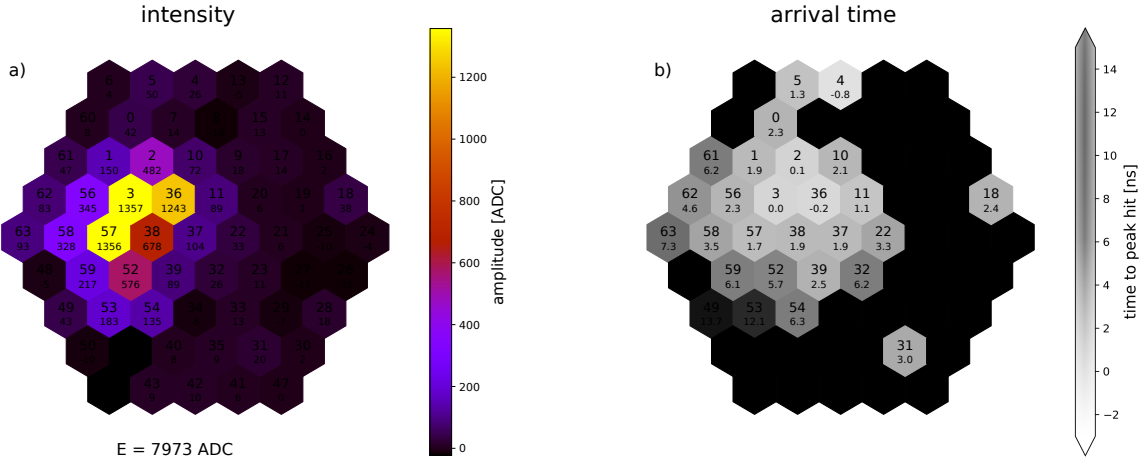}
	\caption{\label{Plt:FirstLightTarget}One of the first events recorded with the IceAct field-telescope. Event amplitude distribution (a) and timing distribution (b) with the typical geometric and time clustering of an air-cherenkov event.}
\end{figure}

Since May 2019 both telescopes are opened and taking data. The roof telescope is synchronized with the IceCube datastream via a trigger output feed into a DOM mainboard, which allows for an easy search for events in coincidence with the IceCube or IceTop datastream. Figure \ref{Plt:FirstLightDRS} shows one of the first events recorded with the roof telescope. One of the first events of the field telescope is shown in figure \ref{Plt:FirstLightTarget}. The typical geometrical clustering (left) and timing (right) of air-cherenkov images is visible. Depending on the configured trigger threshold, the night sky background and weather conditions the telescopes operate between 1\,Hz and 4\,Hz trigger rate with a negligible noise trigger content. 

\begin{figure}
	\centering
	\includegraphics[width=0.8\textwidth]{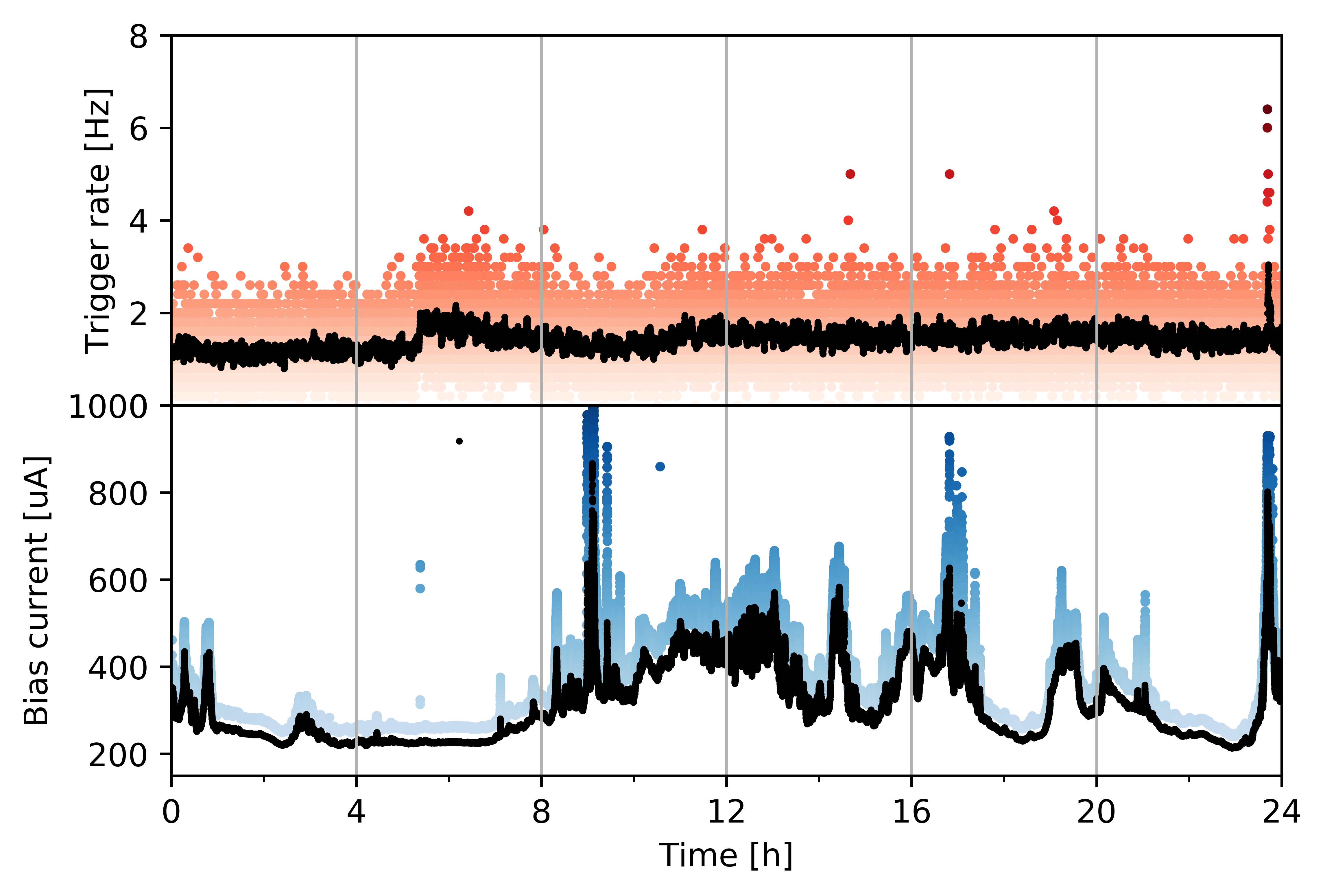}
	\caption{\label{Plt:Moni} Top) Trigger rate (red) and smoothed rate (black) of the IceAct telescope during one day with a threshold of $\approx 54$\,PE per fixed 9 pixel-group (see \cite{Bretz:2018lhg} for further details). Bottom) Single pixel bias current with the maximum pixel current (blue) and the median pixel current (black). Spikes originate from aurora activity in the field-of-view of the telescope. Besides the event around 24:00 these current spikes show little impact on the trigger rate at this trigger level.}
\end{figure}

To evaluate the quality of the data, daily monitoring plots are produced. These monitoring plots include the timeline for the trigger rate and the pixel bias current as well as statistical control plots as the average peak height or the single pixel trigger rate. Figure \ref{Plt:Moni} shows an example trigger rate (Top) and bias current timeline (Bottom) for a full day. The short but strong current peaks originating from aurora activities are clearly visible. The length of the aurora events is in the order of seconds to minutes.  Only very high currents can cause noise high enough to trigger the system: for the day shown only the spike around 24:00 affected the trigger rate significantly. 

\begin{figure}[htb]
	\centering
	\begin{subfigure}[b]{0.49\textwidth}
		\centering
		\includegraphics[width=\textwidth]{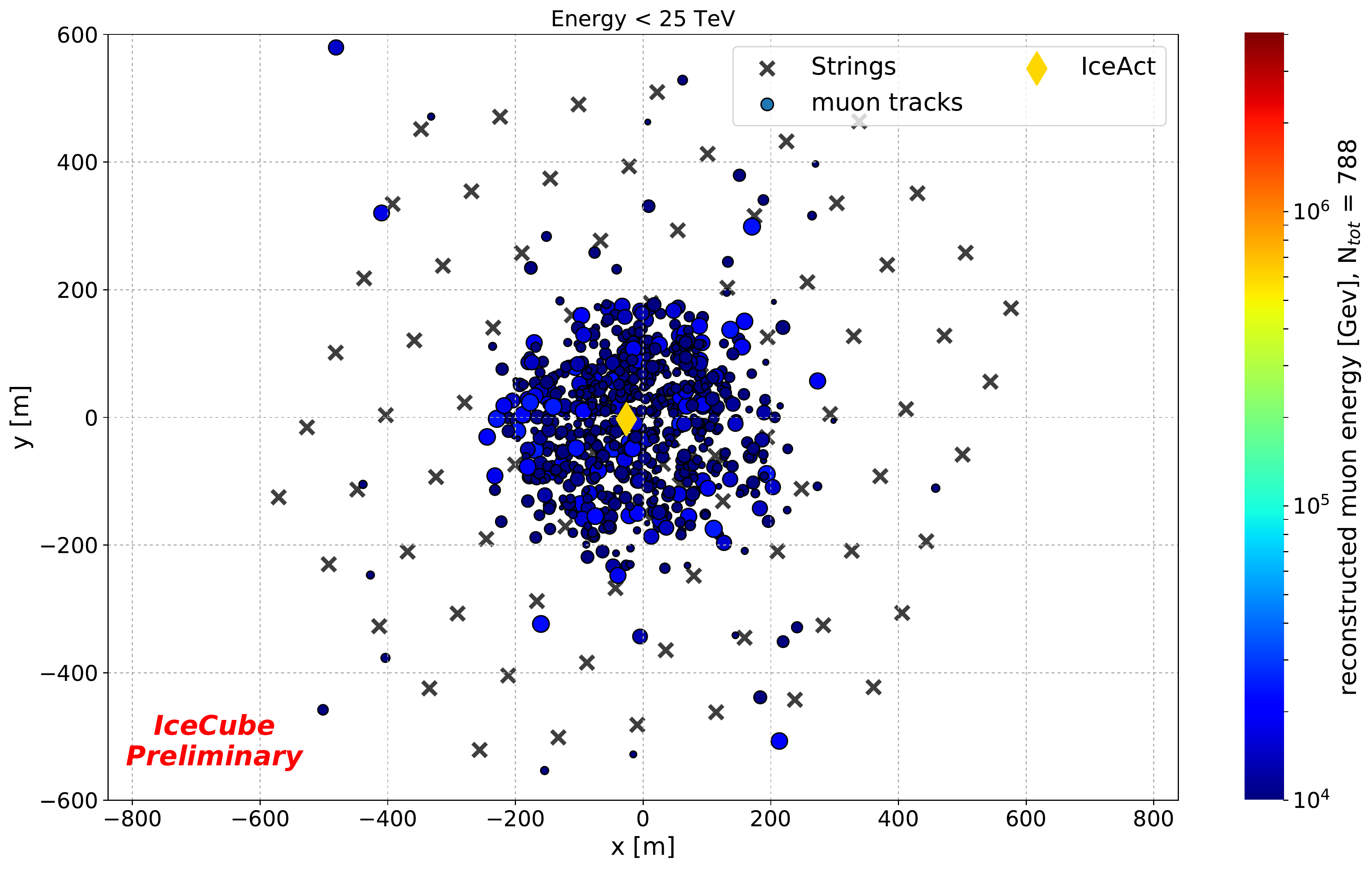}
	\end{subfigure}
	\begin{subfigure}[b]{0.49\textwidth}
		\centering
		\includegraphics[width=\textwidth]{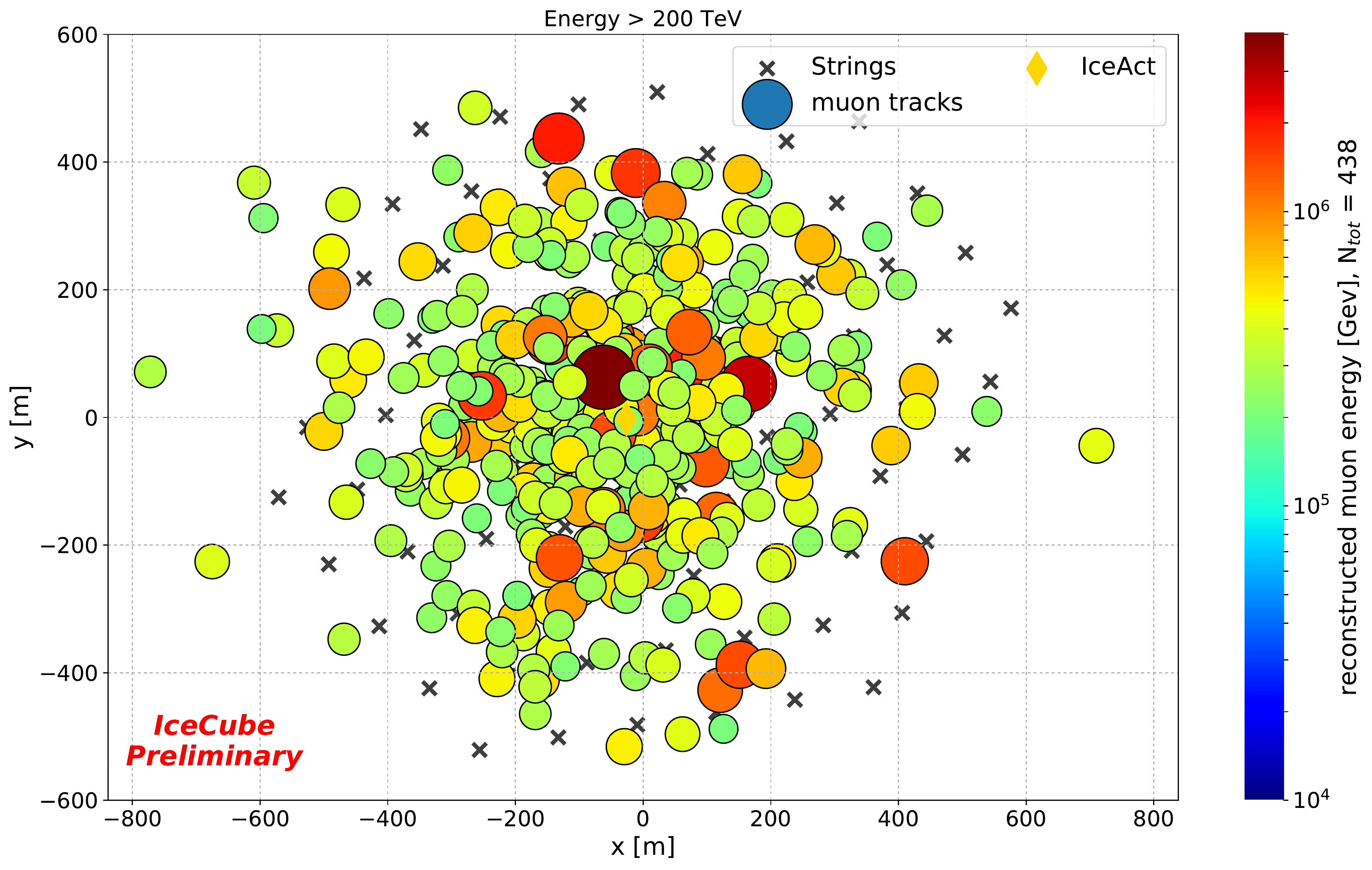}
	\end{subfigure}
	\caption{\label{Plt:Distribution} Coincident events of the IceCube in-ice detector and IceAct with a livetime of one day. Left) Events with a reconstructed energy < 25\,TeV which are clearly clustering around the telescope position. Right) Events with a reconstructed in-ice energy > 200\,TeV. Due to the higher light yield of the high-energy events the effective area increases with event energy.}
\end{figure}

A further cross-check is shown in Figure 
\ref{Plt:Distribution}, which depicts the distribution of reconstructed shower cores on surface for one day of coincident events (triggered by both IceCube and IceAct).  For these distributions, the tracks from high energy muon bundles are detected using the IceCube in-ice detector. The location of the shower core at the surface is then back-tracked using the directional information from IceCube. IceCube's reconstructed muon energy estimation is then used to select events from two different energy regions.  Figure \ref{Plt:Distribution} (left) shows that selecting only the lower energy (< 25~TeV) events: the distribution is clearly centered around the position of IceAct.  Figure \ref{Plt:Distribution} (right) shows coincident events with higher energies (> 200~TeV), which shows that the maximum distance between IceAct and the reconstructed shower cores is clearly correlated with the muon in-ice energy, and that at higher energies the effective area of IceAct is larger for higher energy events. \\
With a detailed study of the instrument using either experimental data (as in \cite{Schaufel:2017niw}) or simulated data, misreconstructed events can be identified.  The energy and angular reconstructions can also be improved by using the image information of the telescope or the hydrid reconstruction with IceTop, thus allowing for more precise event-by-event calibration and cosmic ray studies.

\newpage

\section{IceAct development strategy}

Several stages of telescope arrays are envisioned during the development process that will reach different science goals.
As a first step, a 7 pixel telescope was deployed in 2015 to prove the existence of technology that can be operated at the South Pole.
In 2017 the first telescope with a full 61 pixel camera was operated at the South Pole on the roof of the IceCube Laboratory.
In 2018/2019 this telescope was upgraded and one additional telescope was deployed on the snow surface of the South Pole operating in parallel to other surface detector prototypes. 
The two telescopes currently running at the South Pole test several critical technical capabilities to ensure successful operations of IceAct stations, including: 
\begin{itemize}
\item operation in coincidence with IceCube and IceTop,
\item operation in conjunction with other surface detector prototypes,
\item independent snow management and frost removal, and
\item autonomous response to changing weather conditions and ambient light.
\end{itemize}

In 2019/2020 the two telescopes currently running at the Pole will be tied into the new surface station infrastructure, which has the capability to host up to 7 telescopes.  This will prove the capability of the telescopes to fully synchronize with each other and with IceCube and IceTop based on a White Rabbit timing system. The two telescopes will operate as cosmic-ray detectors in conjunction with IceTop to detect cosmic-ray air showers starting at 30\,TeV up to PeV primary energies.

The 5 additional telescopes to complete the station are planned for deployment in 2020/2021.  This station will have a field of view of $36^{\circ}$ and detect air showers in coincidence with IceTop.  This system will enable composition analyses of cosmic rays at PeV energies with 3 different detector components, IceCube, IceTop, and IceAct, and prove the stable operation of imaging air Cherenkov telescopes under the harsh and variable conditions at the South Pole. It will also test the capability of the stations to act as a veto of cosmic rays for astrophysical neutrino detection.\\   

In the future, a large array of imaging air-Cherenkov telescopes is foreseen. To reach this goal, studies of the capability of the array for gamma-ray detection will also be performed, and a full detector layout will be designed to optimize the capabilities of the different science goals in cosmic-ray, gamma ray and astrophysical neutrino detection.

\section{Summary}

In this proceeding we described the science goals of IceAct, an array consisting of imaging air-Cherenkov telescopes which will operate in conjunction with IceCube, IceTop, and other future astrophysical detector components at the South Pole. In addition, we presented the first light data and the status of the two demonstrator telescopes currently operating at the South Pole.

\newpage
% Set up the bibliography using BibTeX.
% Get references from inspirehep.net or NASA/ADS and put them in references.bib.
\bibliographystyle{ICRC}
\bibliography{references}

% Or, set up the bibliography manually, if you prefer to do things this way.
%
% \begin{thebibliography}{99}
%   \bibitem{Zoll:2015wcu}{{\bf IceCube} Collaboration, \pos{PoS(ICRC2015)1099} (2016).}
%   \bibitem{Peiffer:2017vsm}{{\bf IceCube-Gen2} Collaboration, \pos{PoS(ICRC2017)1052} (2018).}
%   \bibitem{Hussain:2019icrc_gw}{{\bf IceCube} Collaboration, \pos{PoS(ICRC2019)xyz} (these proceedings).}
%   \bibitem{Aartsen:2016nxy}{{\bf IceCube} Collaboration, M.~G.~Aartsen {et al.}, \emph{JINST} {\bf 12} (2017) P03012%
%   % optionally add arXiv ID here [{\tt astro-ph/1612.05093}]
%   .}
%   \bibitem{Waxman:1998yy}{E. Waxman and J. N. Bahcall, \emph{Phys. Rev.} {\bf D59} (1999) 023002.}
% \end{thebibliography}

\end{document}